\documentclass[conference]{IEEEtran}
\IEEEoverridecommandlockouts
\usepackage{cite}
\usepackage{amsmath,amssymb,amsfonts}
\usepackage{algorithmic}
\usepackage{graphicx}
\usepackage{textcomp}
\usepackage{xcolor}

\newcommand*{\nr}     {r}
\newcommand*{\nl}     {l}
\newcommand*{\ns}     {s}
\newcommand*{\xm}     {\vec{\mathbf{x}}_{mean}}

\def\BibTeX{{\rm B\kern-.05em{\sc i\kern-.025em b}\kern-.08em
    T\kern-.1667em\lower.7ex\hbox{E}\kern-.125emX}}
\begin{document}

\title{Dynamic Functional Connectivity Features for Brain State Classification: Insights from the Human Connectome Project
\thanks{The work was supported by the grant for research centers in the field of AI provided by the Ministry of Economic Development of the Russian Federation in accordance with the agreement 000000C313925P4E0002 and the agreement with HSE University № 139-15-2025-009.}
}

\author{
\IEEEauthorblockN{Valeriya Kirova}
\IEEEauthorblockA{\textit{AI and Digital Science Institute} \\ 
\textit{HSE University}\\ Moscow, Russia \\
0000-0001-5333-8425}
\and
\IEEEauthorblockN{Dzerassa Kadieva}
\IEEEauthorblockA{\textit{Institute for Cognitive Neuroscience} \\ 
\textit{HSE University}\\ Moscow, Russia \\
0009-0006-7848-596X}
\and
\IEEEauthorblockN{Daniil Vlasenko}
\IEEEauthorblockA{\textit{Institute for Cognitive Neuroscience} \\ 
\textit{HSE University}\\ Moscow, Russia \\
0009-0002-4867-2896}
\and
\IEEEauthorblockN{Isak B. Blank}
\IEEEauthorblockA{\textit{Institute for Cognitive Neuroscience} \\ 
\textit{HSE University}\\ Moscow, Russia \\
0009-0007-1833-9511}
\and
\IEEEauthorblockN{Fedor Ratnikov}
\IEEEauthorblockA{\textit{AI and Digital Science Institute} \\ 
\textit{HSE University}\\ Moscow, Russia \\
fratnikov@hse.ru, 0000-0003-0762-5583}
}

\maketitle

\begin{abstract}
We analyze functional magnetic resonance imaging (fMRI) data from the Human Connectome Project (HCP) to match brain activities during a range of cognitive tasks. Our findings demonstrate that even basic linear machine learning models can effectively classify brain states and achieve state-of-the-art accuracy, particularly for tasks related to motor functions and language processing. Feature importance ranking allows to identify distinct sets of brain regions whose activation patterns are uniquely associated with specific cognitive functions. These discriminative features provide strong support for the hypothesis of functional specialization across cortical and subcortical areas of the human brain.

Additionally, we investigate the temporal dynamics of the identified brain regions, demonstrating that the time-dependent structure of fMRI signals are essential for shaping functional connectivity between regions: uncorrelated areas are least important for  classification. This temporal perspective provides deeper insights into the formation and modulation of brain neural networks involved in cognitive processing.
\end{abstract}

\begin{IEEEkeywords}
fMRI, brain states, regression, time series
\end{IEEEkeywords}

\section{Introduction}

Modern neuroimaging techniques, such as fMRI, enable the investigation of brain activity in real time, opening new avenues for studying cognitive processes. However, the analysis of fMRI data represents a complex challenge due to its high-dimensional and dynamic nature. In recent years, increasing attention has been paid to the application of network-based data representation methods to describe functional connections between different brain regions\cite{wang2010, richiardi2011, takerkart2014}.

These methods allow for modelling the brain as a network, where nodes represent brain regions and edges represent functional connections between them. This approach enables a deeper investigation into the nature of cognitive states by revealing both local and global connections in brain activity. The use of network-based methods for analyzing fMRI data not only allows for the exploration of characteristics of functional brain networks but also their topology using machine learning (ML) techniques \cite{li2021, saeidi2022, bessadok2023}. We will employ linear ML methods for the classification task. 

The brain contains specific regions and networks that play a key role in classifying and processing various states (emotional, cognitive, perceptual, etc.). It relies on a combination of activity across different regions and their interactions. In this study, we apply ML to identify key brain regions involved in task execution. This research is based on fMRI data from the Human Connectome Project (HCP), specifically the HCP 1200 Subject Release \cite{elam}. The description of the experiments is presented in Section ~\ref{sub53}. This approach will provide a deeper understanding of the brain's functional organization.  

The goals of our research is to demonstrate that classical ML methods effectively handle the classification of fMRI data, even with a limited amount of data  (as shown in Section~\ref{section2}). 
The second goal: to identify key brain regions that are most important for classification and show that each brain state is associated with a unique set of such regions, and to provide an interpretation of the resulting sets  (as demonstrated in Section~\ref{section3}). Identifying the most discriminative neurofunctional features — representing distinct brain regions—is essential for elucidating the relationship between regional neural activity and the cognitive or behavioral states under investigation. These features correspond to brain areas whose activation patterns most strongly contribute to classification accuracy.

Furthermore, we aim to demonstrate that temporal dynamics critically shape functional connectivity between core brain regions, while also identifying region pairs whose interactions remain unaffected by time-dependent variations (see Section~\ref{section4} for detailed discussion).

\section{Data description}

A dataset obtained using fMRI, which describes brain activity during the performance of various cognitive tasks, was utilized. The fMRI data were collected from 581 healthy participants from the HCP. The sample included both men and women aged between 22 and 35 years. Participants performed a series of tasks \cite{ref24}  designed to activate different cortical and subcortical brain networks. For each type of scan, two runs were conducted: one with right-to-left (RL) phase encoding and the other with left-to-right (LR) phase encoding. In total, seven different tasks were performed. Three of these tasks were completed in one session, while the remaining four were performed in another session. 

Each of these tasks was designed to activate specific neural networks, enabling researchers to analyze the functional organization of the brain and its response to various cognitive and emotional stimuli. A brief description of the experiments and the corresponding state labels is provided in the supplementary material.
 
\begin{table}[htbp]
\caption{Two brain states for each cognitive task}
\begin{center}
\begin{tabular}{|c|c|c|}
\hline
\textbf{Cognitive Task} & \textbf{State Name 1} & \textbf{State Name 2} \\
\hline
Working Memory          & 0-back                & 2-back                \\
Gambling                & Win                   & Loss                  \\
Motor Task              & Left hand or foot     & Right hand or foot    \\
Language Processing     & Story                 & Math                  \\
Social Cognition        & Random motion         & Mental interaction    \\
Relational Processing   & Relation              & Similarity            \\
Emotion Processing      & Neutral               & Fear                  \\
\hline
\end{tabular}
\label{task_states}
\end{center}
\end{table}

Thus, we have a total of \( \ns = 14 \) states, each corresponding to a specific type of cognitive task, and \(  8134 \) units of fMRI data.

\textbf{Structure of fMRI data unit.}
We defined the total number of brain regions as  $\nr=379$ . For each brain region, we have a time series of length  $\nl$  describing brain activity over the time interval.   
Thus, each fMRI data unit is represented as an activity matrix  \( \mathbf{X} \) of size $\nr\times \nl$, where each line \( \vec{{x}}_i = (x_{i1}, \dots, x_{i\nl}) \)  corresponds to the activity of the $i$-th brain region over a time interval of length $l$. 
Each element \( x_{ij} \) of the activity matrix \( \mathbf{X} \) corresponds to the activity level of the \( i \)-th brain region at time point \( j \).  

\textbf{Data preprocessing.}
We apply standardization to the data, transforming the values such that their distribution has a mean of 0 and a standard deviation of 1.
We represented each activity matrix \( \mathbf{X} \) by the averaged activity values across brain regions. For each brain region $i=\{1,...,\nr\}$ we calculate the mean value $\overline{\mathbf{x}}_i$ of the time series $\vec{{x}}_i $. Thus, for each activity matrix \( \mathbf{X} \) , we obtain a vector of mean activity values across all brain regions, which will be used in further analysis:  
$$\vec{\mathbf{x}}_{mean}=(\overline{\mathbf{x}}_1,...,\overline{\mathbf{x}}_{\nr}).$$

By focusing on the mean activity values, we aim to demonstrate that each brain state is associated with a unique network of regional activations, which can serve as a reliable signature for classification.

\section{Brain states classification} \label{section2}

Formally, we construct a multi-class classification model  that maps a feature vector $\xm$ , to a set of probabilities for that vector to represent each of $\ns$ brain state classes:   
$$
\xm \rightarrow \begin{pmatrix} 
p_1(\xm)  \\ 
\cdots \\ 
p_{\ns}(\xm)  
\end{pmatrix},
$$
where $p_c(\xm) = p(state = c \mid \xm)$, and \( \ns = 14 \) classes again correspond to distinct brain states. 

We begin with the simplest ML approach -- linear models -- to classify brain states, motivated by several key factors related to both our data characteristics and analytical goals. Their interpretability allows direct examination of weights highlighting important brain regions, while their efficiency with moderate-sized fMRI datasets helps prevent overfitting. These models demonstrate particular robustness to noise, maintaining reliable performance despite common artifacts in neuroimaging data. Their computational simplicity enables rapid processing of high-dimensional neural data without sacrificing predictive accuracy. Furthermore, linear models provide a flexible foundation that can be extended to capture more complex relationships when needed. This approach aligns well with established neuroscientific theory suggesting that cognitive processes emerge from linear combinations of regional brain activations.

We conducted a comprehensive analysis of linear ML models for multiclass classification of brain states using fMRI data. Our evaluation revealed that all linear models achieved high classification accuracy, with most models clustering around $0.9$ accuracy as shown in Table \ref{tab2}. These results demonstrate that linear approaches retain competitive performance for brain state classification, suggesting their continued relevance in fMRI decoding tasks despite advances in nonlinear methods.

We chose logistic regression with a One-vs-All (OvA) strategy to achieve the main goals of our research. 
This choice is motivated by the importance of being able to clearly distinguish each brain state from all others during classification. Technically, in the OvA approach, multiclass classification is performed by training $\ns$ independent binary classifiers, one for each brain state class $ c=\{1,...,\ns\} $. 
Each such classifier is trained to distinguish one brain state class $c$ from all others by computing a corresponding weight vector $\vec{\mathbf{w}}_c = (w_1^{}, \dots, w_{\nr})$. This vector represents the contribution of each feature — corresponding to the mean activity of a specific brain region —  to the identification of brain state $c$. 
This directly enables us to analyse the contribution of individual brain regions to the discrimination of each specific state from all others, which is a key requirement for neurobiological interpretation.



\begin{table}[htbp]
\caption{Accuracy of linear classification models}
\begin{center}
\begin{tabular}{|c|c|}
\hline
\textbf{Model} & \textbf{Accuracy} \\ 
\hline
    Passive-Aggressive Classifier & 0.84 \\
    SGD Classifier with Log Loss & 0.85 \\
    Perceptron & 0.85 \\
    Linear SVM (OvR) & 0.85 \\
    Linear SVM (LibLinear) & 0.88 \\
    L1-Logistic Regression & 0.89 \\
    L2-Logistic Regression & 0.89 \\
    One-vs-All Linear Regression & 0.90 \\
    Multinomial Logistic Regression & 0.90 \\
    RBF Kernel SVM (OvR) & 0.90 \\
    Ridge Classifier & 0.90 \\
    Linear Discriminant Analysis & 0.92 \\
\hline
\end{tabular}
\label{tab2}
\end{center}
\end{table}

\begin{table}[htbp]
  \caption{Classification report with OvA}
  \label{tab:classification-report}
\begin{center}
\begin{tabular}{|c|c|c|}
\hline
    \textbf{Brain state} & \textbf{Accuracy} & \textbf{Samples} \\
\hline
    Neutral       & 0.88 & 120  \\
    Fear          & 0.87 & 124  \\
    Loss          & 0.78 & 108  \\
    Win           & 0.85 & 117  \\
    Math          & 0.97 & 120  \\
    Story         & 0.96 & 104  \\
    Random motion & 0.97 & 125  \\
    Mental interaction & 0.95 & 99  \\
    0-back        & 0.83 & 127  \\
    2-back        & 0.78 & 98   \\
    Left hand/foot & 0.92 & 129 \\
    Right hand/foot & 0.99 & 116 \\
    Similarity    & 0.79 & 124  \\
    Relation      & 0.88 & 116  \\
\hline
\end{tabular}
\end{center}
\end{table}

The classification results for each brain state class are summarised in Table~\ref{tab:classification-report}. 
For each brain state class, we report both the class-specific accuracy and the actual number of test samples.
The total test set size is $1627$ measurements. The model's Accuracy of $0.9$ (from Table ~\ref{tab2}) indicates it correctly predicted $90\%$, meaning approximately $1464$ correct predictions.

We highlight the high precision of the states —
 \emph{left hand or foot, and right hand or foot,  math, story, mental interaction, random motion} — which correspond to experiments in \textbf{Motor Task} for mapping brain motor regions, \textbf{Language Processing} for studying brain areas involved in language comprehension and arithmetic processing, and \textbf{Social Cognition} for identifying neural mechanisms of social interaction and intention recognition.

\section{Significant brain regions identification} \label{section3}

To identify the most significant brain regions that contribute the most to the classification of different brain states, we employ the following approach.

\subsection{Analysis of model accuracy dependency on the number of used brain regions}

For each brain state class $ c=\{1,...,\ns\} $, we rank the features -- averaged brain regions signals $\vec{\mathbf{x}}_{mean}$ ,  according to the absolute values of their corresponding weights in  $\vec{\mathbf{w}}_c = (w_1^{}, \dots, w_{\nr})$. This procedure established a distinct feature importance ranking for each brain state class. 

\begin{figure}[tb]
\centerline{\includegraphics[width=\linewidth]{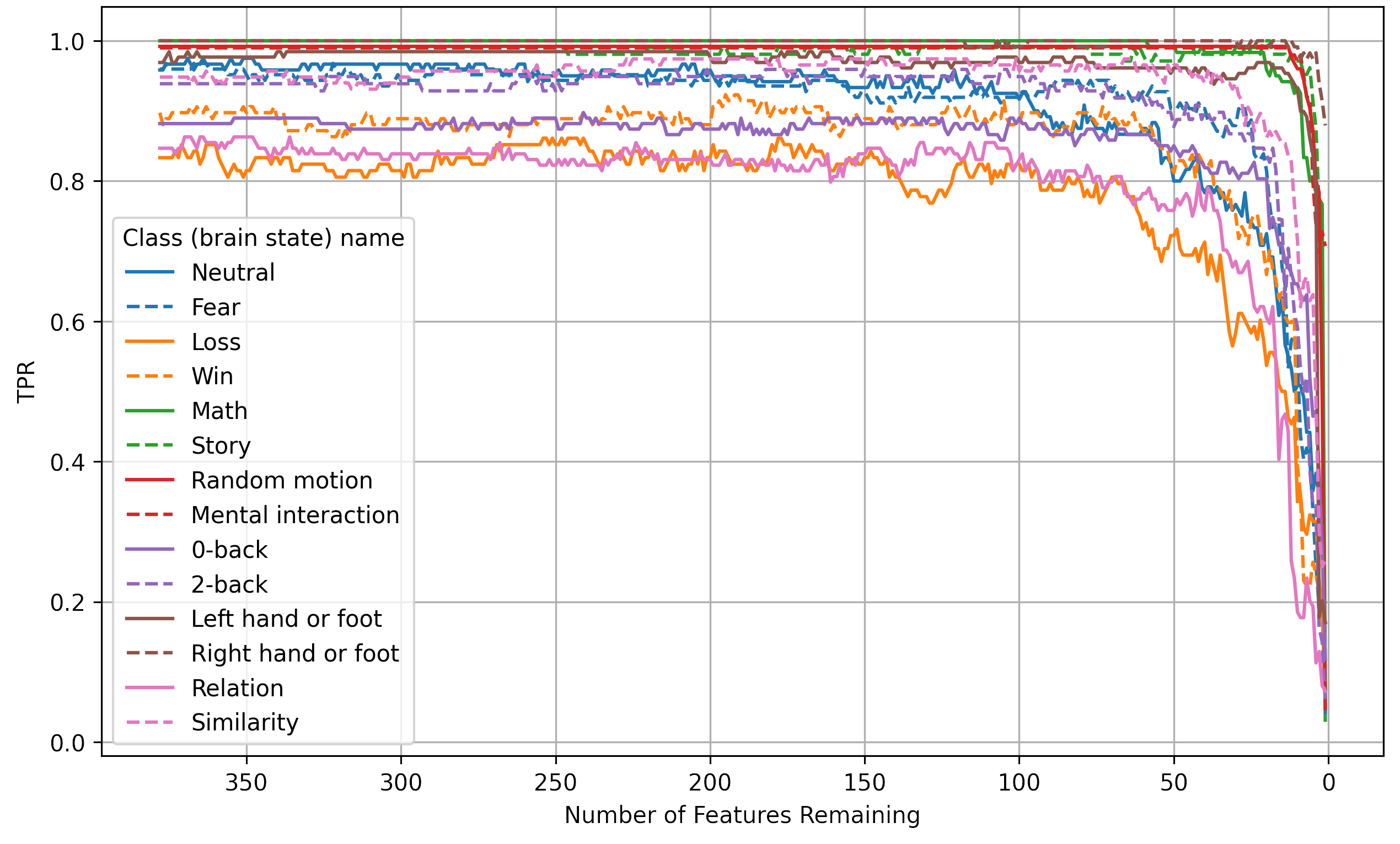}}
\caption{Accuracy plot depending on the number of features.}
\label{fig:accuracy-plot}
\end{figure}

To analyse the dependence of model accuracy on the number of used features, we applied the following approach: we iteratively removed features, starting from the least significant ones. 
At each iteration, we retrained the model on the reduced feature set,  re-ranked the remaining features by their updated absolute weights, and evaluated per-class performance using the True Positive Rate (TPR or Recall). We chose TPR because it directly measures how feature reduction impacts the model’s ability to detect all instances of each class. As TPR quantifies the proportion of target examples correctly identified, it is critical for our task where missing true positives is unacceptable. The TPR values (averaged across classes) were recorded and visualized on the Fig. \ref{fig:accuracy-plot}, with the $X$-axis showing the number of remaining features and the $Y$-axis showing the corresponding accuracy. 

This graph clearly demonstrates that feature selection can be performed without significant loss of accuracy up to a certain limit. However, as the number of features is further reduced, the model begins to lose its ability to classify adequately, highlighting the importance of selecting the right number of features to maintain a balance between model simplicity and its quality.

\subsection{Significant brain regions selection}

We begin with the complete set of significant brain regions and iteratively remove features in descending order of importance, as determined by absolute weight values. After each removal step, we: (1) evaluate the model's true positive rate (TPR), and (2) re-rank the remaining features based on their updated weights in the refined model. 
If the TPR drops by more than a specified threshold   ( we tested  for  $5\%$  and $10\%$) relative to the maximum value observed prior to feature elimination, the last removed feature is restored and the procedure terminates. This criterion ensures that feature elimination continues only while the TPR remains within acceptable limits.
This method allows us to retain key brain regions that significantly impact the model's TPR while eliminating less significant features, minimizing the loss in classification quality. 

As a result, for each brain state class $c\in \{1,...,s\}$, we obtain an informative set of significant brain regions:
$$\mathcal{M}_c=\{i ~|~ i \text{ is the index of significant brain region  for class } c\},$$ 
which will be further analyzed.

\begin{figure}[tb]
\centerline{\includegraphics[width=\linewidth]{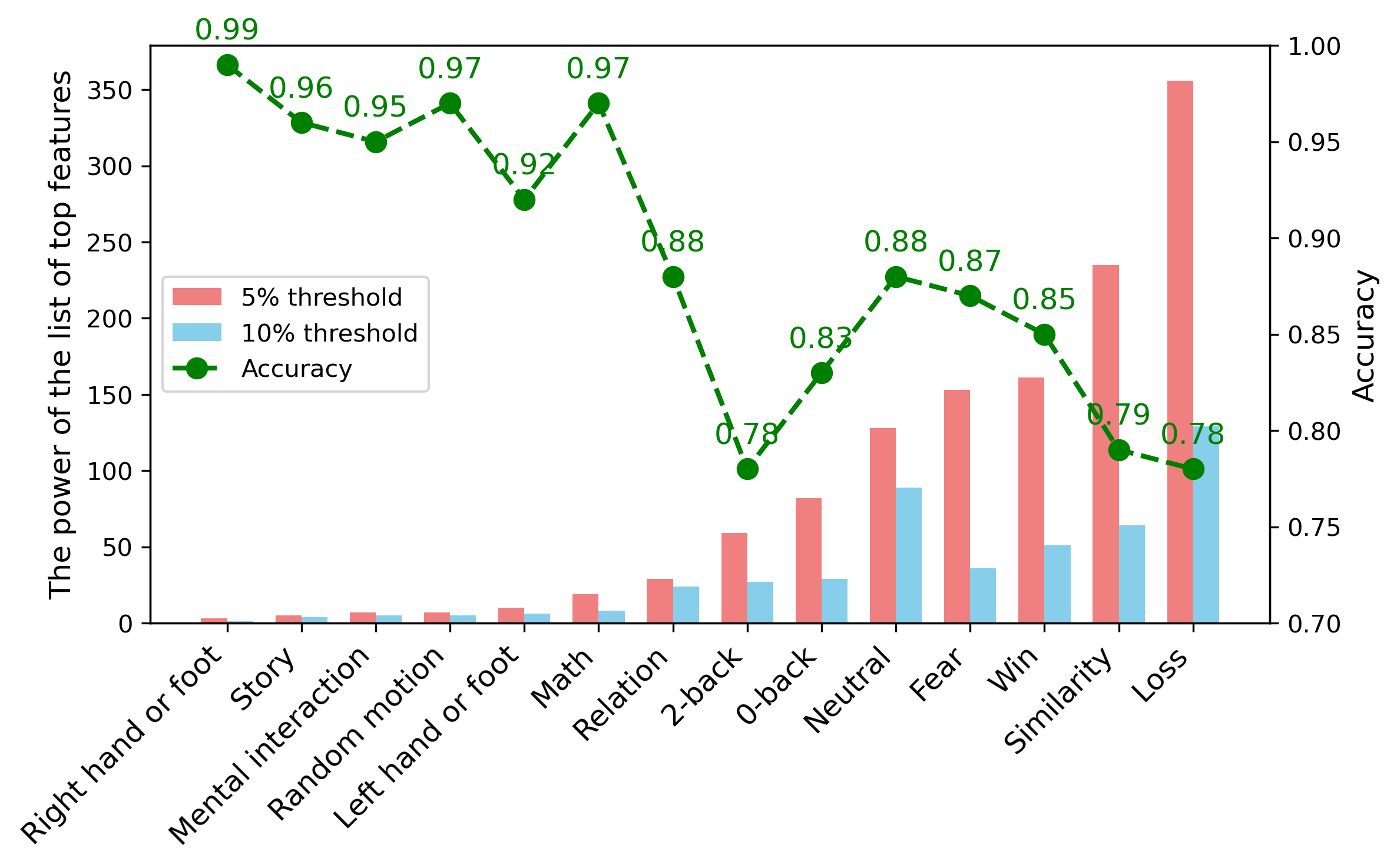}}
\caption{The graphs of cardinality of the top features sets.}
\label{Fig1}
\end{figure}

Fig.~\ref{Fig1} presents the distribution of significant brain regions set sizes across different classes at accuracy reduction thresholds  of $5\%$ and $10\%$. The results demonstrate  substantial inter-class variability in the number of brain regions required for effective classification.  

Classes exhibiting high classification accuracy consistently show compact sets of significant brain regions at both thresholds, indicating that these brain states are characterized by specific, localized neural activity patterns that facilitate robust discrimination. 
Conversely, states with lower classification accuracy require substantially larger feature brain regions for characterization. This pattern holds true at both examined thresholds, suggesting that these brain states lack clearly identifiable focal signatures. Instead, their neural representations appear to involve distributed activity across multiple cortical areas, necessitating integration of information from broader neural networks for accurate classification. 

\subsection{Uniqueness of significant brain regions sets}

Our analysis will utilize feature sets generated at a 5\% accuracy drop threshold. 
To quantify the similarity between the sets of significant brain regions corresponding to different brain state classes, we compute the Jaccard coefficient for each pair of classes $i$ and $j$.  The Jaccard coefficient is computed as the ratio of the intersection of sets to the union of sets:
\[
J(\mathcal{M}_i, \mathcal{M}_j) = \frac{|\mathcal{M}_i \cap \mathcal{M}_j|}{|\mathcal{M}_i \cup \mathcal{M}_j|}, \quad i,j \in \{1, 2, \dots, \ns\}.
\]
The Jaccard coefficient ranges between \( [0, 1] \), where $1$ indicates that the sets are identical,   $0$ indicates that the sets have no common elements.
By pairwise comparing the sets of significant brain regions, we construct a Jaccard coefficient matrix:
\[
\mathbf{J} = \{J(\mathcal{M}_i, \mathcal{M}_j)\}, \quad i,j \in \{1, 2, \dots, \ns\}.
\]
The Jaccard coefficient matrix \( \mathbf{J} \) is a symmetric matrix of size \( \ns \times \ns \). This matrix allows us to assess the degree of overlap between significant brain regions across different brain state classes.

\begin{figure}[tb]
\centerline{\includegraphics[width=\linewidth]{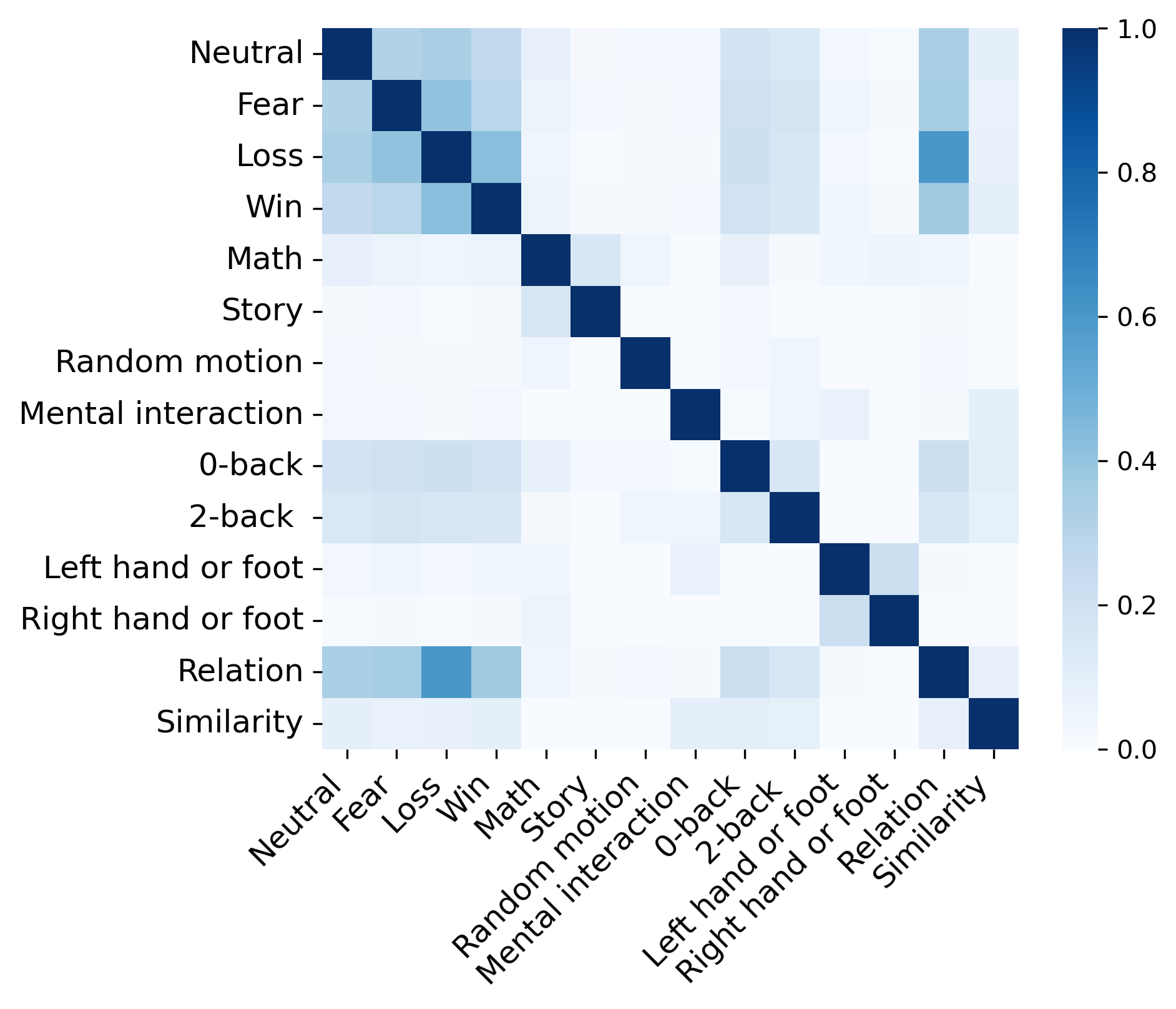}}
\caption{The Jaccard coefficient matrix \( \mathbf{J} \) heatmap.}
\label{Fig3}
\end{figure}

Furthermore, Fig.~\ref{Fig3} shows that the sets of significant brain regions \( \mathcal{M}_i \), corresponding to brain states with high classification accuracy, hardly overlap. This indicates that for each brain state, the model has identified a unique set of regions whose mean activity is most significantly associated with that state. However, there are also sets whose elements are similar in composition. This occurs when these sets approach the comprehensive coverage of the HCP MMP atlas ($\nr=379$ regions), resulting in reduced specificity of significant region identification. Under these conditions, the distinctive contributions of individual regions become less discernible, complicating the interpretation of their functional roles in brain state classification. 

The proposed approach not only allows us to identify the most significant brain regions for each condition but also reveals the relationship between the number of significant brain regions and classification accuracy. For states with high classification accuracy, distinct activation patterns emerge in specific brain regions, clearly delineating the neural signatures associated with each condition.  This highlights the importance of analysing the spatial structure of brain activation to understand the neurobiological mechanisms underlying different states.


\subsection{Analysis of the importance of temporal dynamics of significant brain regions} \label{section4}

For the classes with high classification accuracy — \emph{math, story, random motion, mental interaction, left hand or foot, and right hand or foot} — our sequential analysis focuses on these six brain conditions. 
We focus on such classes because the small number of significant brain regions indicates the presence of clearly defined and localized patterns of brain activation that play a key role in the formation of these states. 

One of the main tasks of our analysis is to verify the significance of the temporal structure in identifying functional connections between brain regions. We work with fMRI data, where each significant region is represented by a time series. By studying the correlations between these time series, we aim to determine whether temporal dynamics significantly contribute to the formation of functional connections or whether these connections can be identified based on averaged or static characteristics of activity.

\begin{figure}[tb]
\centerline{\includegraphics[width=\linewidth]{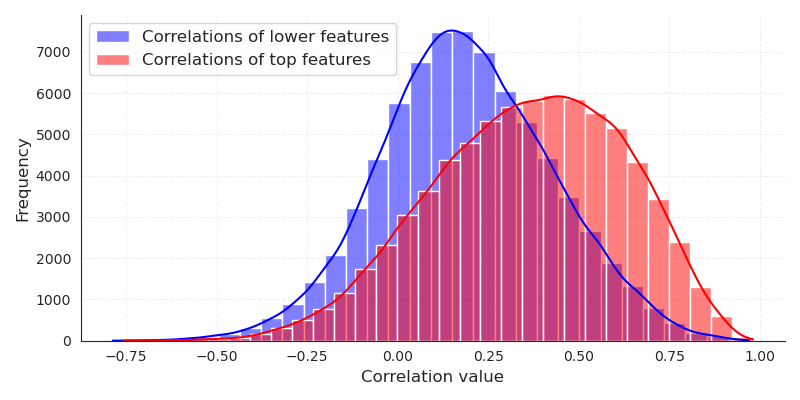}}
\caption{Distributions of time series correlation of top and least features.}
\label{Fig5}
\end{figure}

We analyze the correlations between the time series of highly informative brain regions, selected based on their contribution to the classification model. We compute the distribution of pairwise correlations among these regions to assess the degree of their synchronized dynamics.
Similarly, we select low-weight regions — in the same number as the top features — and construct the correlation distribution of their time series. Fig. \ref{Fig5} shows both distributions: the first corresponds to the most informative features, while the second shows the least informative ones.

To quantitatively assess the difference between these distributions, we treat them as representing two distinct classes: class 0 for the top features and class 1 for the low-weight features. We then construct an ROC curve by comparing TPR and FPR across varying thresholds. The resulting AUC = 0.7 confirms that the correlation distribution of top features offers stronger class separation.

To analyze in more depth the importance of temporal structure in functional connections between significant brain regions, we perform the following steps, as described in sections \ref{sub51}, \ref{sub52}, \ref{sub53}.

\subsection{Constructing Distributions of Correlations of Time Series of significant brain regions}
\label{sub51}
For each activity matrix \( \mathbf{X} \) corresponding to brain state class $c$, we perform the following steps.

1. We extract the time series \(\{\overline{{x}}_{i_1}, \overline{{x}}_{i_2}, \dots, \overline{{x}}_{i_{t}}\}\) for significant brain regions, where the region indices belong to the set 
   \[
   \mathcal{M}_c = \{i_1, i_2, \dots, i_{t}\}.
   \]

2.  We construct the correlation matrix  $\text{corr}(\overline{{x}}_{i}, \overline{{x}}_{j}),$ where \( i, j \in \mathcal{M}_c \) are the indices of the selected significant brain regions of the extracted time series, where each element represents the Pearson correlation coefficient between two time series.

Constructing correlation matrices of time series allows us to study the relationship between the activities of different brain regions that have been identified as significant for each state \( c \). Pearson correlation coefficients show how closely the activity of one region changes relative to another within a single observation. High correlation values may indicate synchronous activation of regions, which could suggest their functional relationship in the context of the given brain state.

For each pair $(i,j)$ of significant brain regions indices $i, j \in \mathcal{M}_c$  we construct
\[ P_c(\text{corr}(\overline{{x}}_{i}, \overline{{x}}_{j})) \] --
the distributions of the Pearson correlation coefficients between time series of significant brain regions $i,j\in \mathcal{M}_c$ for all activity matrices of class $c$.

Analysing the distributions allows us to assess the stability and nature of interactions between pairs of brain regions within a single state. For example, if the distribution of correlation coefficients between two regions shows a consistently high value, this may indicate their systematic co-activation. In contrast, a wide or sparse distribution may suggest variability in the connection between these regions.

\subsection{Constructing distributions of correlations of shuffled time series of significant brain regions}
\label{sub52}

In the original data, the time series of significant brain regions  \(\{\overline{{x}}_{i_1}, \overline{{x}}_{i_2}, \dots, \overline{{x}}_{i_{t}}\}\) reflect the dynamic activity of specific brain regions, and their temporal dependencies play an important role in the formation of functional patterns and interactions between regions.  
To what extent is the temporal structure significant in identifying functional connections between brain regions? To answer this question, we perform the following steps.

In each extracted subset  of time series \(\{\overline{{x}}_{i_1}, \overline{{x}}_{i_2}, \dots, \overline{{x}}_{i_{t}}\}\), we shuffle all time series to create a control dataset where the temporal structure of brain region activity is completely disrupted, and the original dependencies within the time series are eliminated. Critically, this shuffling preserves all statistical properties of the individual time series—such as mean, variance, and higher-order moments—while only destroying the temporal dynamics and dependencies between time points.

To denote the shuffled values of the time series we introduce the following notation:
\[
\overline{{x}}_i ^{{\text{shuffled}}} = (x_{\pi(1)}, \dots, x_{\pi(\nl)}),
\]
where \( \pi: \{1, 2, \dots, \nl\} \to \{1, 2, \dots, \nl \} \) is a random permutation of the time point indices.

In the shuffled  series there is no temporal structure, which eliminates the influence of dynamic coordination between brain regions.

Similarly to the previous section, we construct 
\[P^{\text{shuffled}}(\text{corr}(\overline{{x}}_{i}^{\text{shuffled}} , \overline{{x}}_{j}^{\text{shuffled}} ))\] — the distribution  of the Pearson correlation coefficients between the shuffled time series of significant brain regions $i,j\in \mathcal{M}_c$ for all activity matrices of class $c$.

\subsection{Analysis of the significance of temporal structure}
\label{sub53}
We compare the distributions of original and shuffled time series to determine which properties of temporal activity are truly critical for the formation of functional patterns associated with brain states. 
To analyse the differences between distributions and assess the influence of temporal structure on correlation coefficients, we perform the following analysis for each significant brain regions pair \( (i, j) \) where \( i, j \in \mathcal{M}_c \). We conduct a Kolmogorov-Smirnov (KS) test to compare: the distribution of Pearson correlation coefficients from original time series of significant brain regions, versus the distribution from shuffled time series. Formally, the KS test examines the hypothesis that the two distributions  originate from the same underlying population.

The KS test statistic is defined as:
\[
D_{i,j} = \sup_r \left| F(r) - F^{\text{shuffled}}(r)\right|,\]
where \[F(r) = P(\text{corr}(\overline{x}_i, \overline{x}_j) \leq r)\] is the correlation distribution function of the original time series,  
\[ F^{\text{shuffled}}(r) = P\big(\text{corr}(\overline{x}_i^{\text{shuffled}}, \overline{x}_j^{\text{shuffled}}) \leq r\big) \] is the correlation distribution function of the shuffled time series.

Based on the value of the test statistic \( D_{i, j} \) and the corresponding \( p \)-value, we can assess the significance of the differences between the distributions. We set the significance level at \( \alpha = 0.03 \). If the \( p \)-value is less than \( \alpha \), the null hypothesis of equal distributions is rejected, indicating significant differences in the correlation structure between the original and shuffled time series for a given pair of features \( i \) and \( j \) in class \( c \).

Thus, for each pair of features \( i \) and \( j \) within class \( c \), we obtain the KS statistic \( D_{i, j} \) and the corresponding \( p \)-value, allowing us to evaluate the significance of differences in the correlation structure.

The KS test results demonstrate that for nearly all feature pairs  \( (i, j) \) in each class, the \( p \)-values were significantly below the chosen significance threshold of \( \alpha = 0.03 \).
Upon evaluating all significant regions associated with high-accuracy classes, none of the feature pairs showed   \(p\)-values above the predefined significance level  \( \alpha  \).
This confirms that the distributions of correlation coefficients from original time series are statistically distinct from those derived from shuffled time series. 
This suggests that the temporal structure of the data plays a crucial role in shaping the functional connections between brain regions. Disrupting the temporal structure by shuffling the data leads to substantial changes in correlation, confirming the hypothesis that the temporal dynamics of brain activity are a key factor in forming functional interactions between brain regions.

\section{Conclusion}

We analyzed fMRI data reflecting brain activity during various cognitive tasks. By averaging the activation values for each brain region, we applied machine learning methods to demonstrate that each cognitive state is characterized by a unique pattern of brain activity.

Our work highlights the relevance and applicability of classical machine learning algorithms for fMRI data analysis, especially in scenarios with limited data and a strong need for interpretability. All classical classification methods demonstrated high performance, particularly for states related to motor functions and speech processing. For some categories, accuracy reached nearly $100\%$, indicating highly distinctive and expressive brain activity patterns associated with these states.

A key part of the study was identifying brain regions whose activity was most closely associated with specific tasks. For each cognitive state, we determined a unique set of features representing brain areas significantly involved in the corresponding type of cognitive activity. An analysis of model accuracy depending on the number of features used showed that high classification accuracy often required only a relatively small number of significant features. This suggests the presence of specific brain regions that are crucial for forming these states, enabling accurate classification.

Further analysis using the Jaccard coefficient confirmed that feature sets for highly classifiable states were unique, emphasizing their specificity and functional uniqueness. These results support the hypothesis that each brain state is characterized by its own distinctive pattern of activity shaped by specialized neural network processes.

Moreover, the set of most significant features aligns with existing neuroscientific research on brain functions, confirming that our findings correspond to regions associated with each state in the literature. This demonstrates that our machine learning-based approach effectively identifies neural correlates of cognitive states, confirming the robustness of our methodology. Therefore, our study not only reinforces existing knowledge through data-driven analysis but also provides a solid foundation for further exploration of complex brain activation patterns.

For the states with high classification accuracy, we demonstrated the importance of temporal dynamics in brain activity for establishing functional connectivity. We showed that brain regions with a high contribution to classification exhibited stronger temporal correlations than those with minimal contribution. To further analyze the role of temporal structure in these correlations, we disrupted the temporal dynamics of the time series and computed correlation distributions for the shuffled data.

A Kolmogorov–Smirnov test confirmed significant differences between the correlation coefficient distributions of the original and shuffled time series, with 
$p$-values close to zero for most feature pairs. This indicates that disrupting the temporal structure leads to substantial changes in correlation patterns, supporting the hypothesis that the temporal dynamics of brain activity play a key role in shaping functional interactions between brain regions.

The obtained results can be utilized for further exploration of the neurobiological mechanisms underlying cognitive processes. The proposed approach enables effective analysis of both spatial and temporal dynamics of brain activity, identifying significant regions and their interactions. This makes it a valuable tool for studying the neural basis of cognitive processes and the functional organization of the brain, especially in scenarios with limited data.

\section{Supplementary Material} \label{supp}

Description of tasks completed in the fMRI scanner.

\textbf{Working Memory.} This task was designed to study short-term memory and information retention processes. Participants were shown blocks of images depicting places, tools, faces, and body parts. Each run included blocks from four types of stimuli, with half of the blocks performed with a 2-back task requiring memory retention, and the other half with a 0-back task for comparison. The 2-back task required participants to remember the sequence of images and determine whether the current image matched the one shown two steps earlier. The 0-back task only required participants to identify whether the current image matched a given target. Task blocks alternated with fixation blocks, where participants looked at a cross on the screen, allowing researchers to track brain activity changes in the absence of cognitive load.

\textbf{Gambling Task:} This task simulated decision-making processes under uncertainty and risk. Participants played a card game in which they had to guess whether the number on a hidden card was higher or lower than 5. Depending on their answer, they could win (green upward arrow with "1"), lose (red downward arrow with "0.50"), or receive no reward (neutral outcome). The task was divided into blocks dominated by either winning or losing outcomes, enabling researchers to examine how the brain responds to reward anticipation and loss.

\textbf{Motor Task.} This task was aimed at mapping motor areas of the brain. Participants were presented with visual cues instructing them to perform movements such as tapping their fingers on the left or right hand, squeezing their toes, or moving their tongue. Movement blocks lasted 12 seconds, and each of the two runs contained 13 blocks, including tongue, hand, and foot movements, as well as fixation blocks. These tasks help identify motor cortex areas activated in response to different types of movements.

\textbf{Language Processing.} This task aimed to study language comprehension and arithmetic processing. In the story blocks, participants listened to short narratives, such as adaptations of Aesop’s fables, and then answered questions about the story content. In the math blocks, participants solved verbal arithmetic problems. The tasks alternated, and the block durations were adjusted to ensure equal completion times, allowing comparison of brain activity during language processing and numerical operations.

\textbf{Social Cognition (Theory of Mind).} Participants were shown short video clips where geometric shapes (squares, circles, triangles) either interacted with each other or moved randomly. After each video, participants evaluated whether the shapes had intentions and were interacting or whether their movements were random. This task allowed researchers to study brain areas associated with understanding social interactions and recognizing intentions.
 
\textbf{Relational Processing.} This task involved analyzing relationships between objects on the screen. In one condition, participants determined which characteristic (shape or texture) differentiated object pairs and assessed whether this distinction applied to another pair. In the control condition, they simply identified whether the lower object matched one of the upper objects based on a given criterion. This task helped researchers study the brain’s ability to perform comparative analysis and identify relationships between objects.

\textbf{Emotion Processing.} Participants were shown faces displaying fear or anger, as well as neutral figures, and had to choose one of two faces or figures that matched the presented stimulus. Task blocks alternated with fixation blocks, allowing researchers to assess brain activity during emotional stimulus processing compared to neutral conditions.

\section*{Acknowledgment}
This research is supported in part through computational resources of HPC facilities at HSE University.


\end{document}